\documentclass[aps,prl,twocolumn,preprintnumbers,superscriptaddress,nofootinbib,floatfix]{revtex4-2}

\usepackage{graphicx,epstopdf}
\usepackage{amssymb,amsmath}
\usepackage{dcolumn}
\usepackage{bm}
\usepackage{color}
\usepackage{enumerate}
\usepackage[colorlinks=true,citecolor=blue]{hyperref}
\hypersetup{colorlinks=true,citecolor=blue,linkcolor=blue,urlcolor=blue}
\usepackage{soul}

\DeclareMathAlphabet\mathbfcal{OMS}{cmsy}{b}{n}

\usepackage{tabularx}
\usepackage{comment}

\begin{document}

\title{Exciton Spin Hall Effect In Arc-Shaped Strained WSe$_2$}

\author{A. Shubnic}
\affiliation{Abrikosov Center for Theoretical Physics, MIPT, Dolgoprudnyi, Moscow Region 141701, Russia}
\affiliation{Department of Physics, ITMO University, Saint Petersburg 197101, Russia}

\author{V. Shahnazaryan}
\email{vanikshahnazaryan@gmail.com}
\affiliation{Abrikosov Center for Theoretical Physics, MIPT, Dolgoprudnyi, Moscow Region 141701, Russia}
\affiliation{Department of Physics, ITMO University, Saint Petersburg 197101, Russia}

\author{I. A. Shelykh}
\affiliation{Science Institute, University of Iceland, Dunhagi 3, IS-107, Reykjavik, Iceland}
\affiliation{Abrikosov Center for Theoretical Physics, MIPT, Dolgoprudnyi, Moscow Region 141701, Russia}
\affiliation{Russian Quantum Center, Skolkovo IC, Bolshoy Bulvar 30 bld. 1, Moscow 121205, Russia}

\author{H. Rostami} 
\email{hr745@bath.ac.uk}
\thanks{Current affiliation of H. Rostami is: Department of Physics, University of Bath, Claverton Down, Bath BA2 7AY, United Kingdom}

\affiliation{Nordita, KTH Royal Institute of Technology and Stockholm University, Hannes Alfvéns väg 12, 10691 Stockholm, Sweden}

\begin{abstract}
Generating a pure spin current using electrons, which have degrees of freedom beyond spin, such as electric charge and valley index, presents challenges.
In response, we propose a novel mechanism based on intervalley exciton dynamics in {\em arc-shaped} strained transition metal dichalcogenides (TMDs) to achieve the {\em exciton spin Hall effect} in an electrically insulating regime, without the need for an external electric field.
The interplay between strain gradients and strain-induced pseudomagnetic fields results in a net Lorentz force on long-lived intervalley excitons in WSe$_2$, carrying non-zero spin angular momentum.
This process generates an exciton-mediated pure spin Hall current, resulting in opposite-sign spin accumulations and local magnetization on the two sides of the single-layer arc-shaped TMD.
We demonstrate that the magnetic field induced by spin accumulation, at approximately $\sim {\rm mT}$, can be detected using techniques such as superconducting quantum interference magnetometry or spatially-resolved magneto-optical Faraday and Kerr rotations.
\end{abstract}
\date{\today}

\maketitle

{\em Introduction.} The spin Hall effect (SHE) was first introduced by D'yakonov, Perel' in 1971 \cite{Dyakonov_Perel_1971}, revisited by Hirsch in 1999 \cite{Hirsch1999} and experimentally demonstrated in 2004 \cite{Kato2004,Wunderlich_PRL_2005}. One of the versions of SHE, namely the intrinsic SHE arises from strong spin-orbit coupling resulting in the band structure quantum geometry (Berry curvature), and leading to the appearance of a transverse spin current when an electric field is applied, without the need of an external magnetic field. This effect generates edge spin accumulation, observed in GaAs-based semiconductors using magneto-optical Kerr rotation measurement \cite{Kato2004}. The quantum spin Hall effect predicted in graphene \cite{Kane2005} and 2D electron gases \cite{Bernevig2006} was experimentally verified in HgTe, revealing quantized spin-Hall conductance and absent charge-Hall conductance \cite{Bernevig_Science2006}.

\begin{figure}[h!]
    \centering
    \includegraphics[width = 0.75\linewidth]{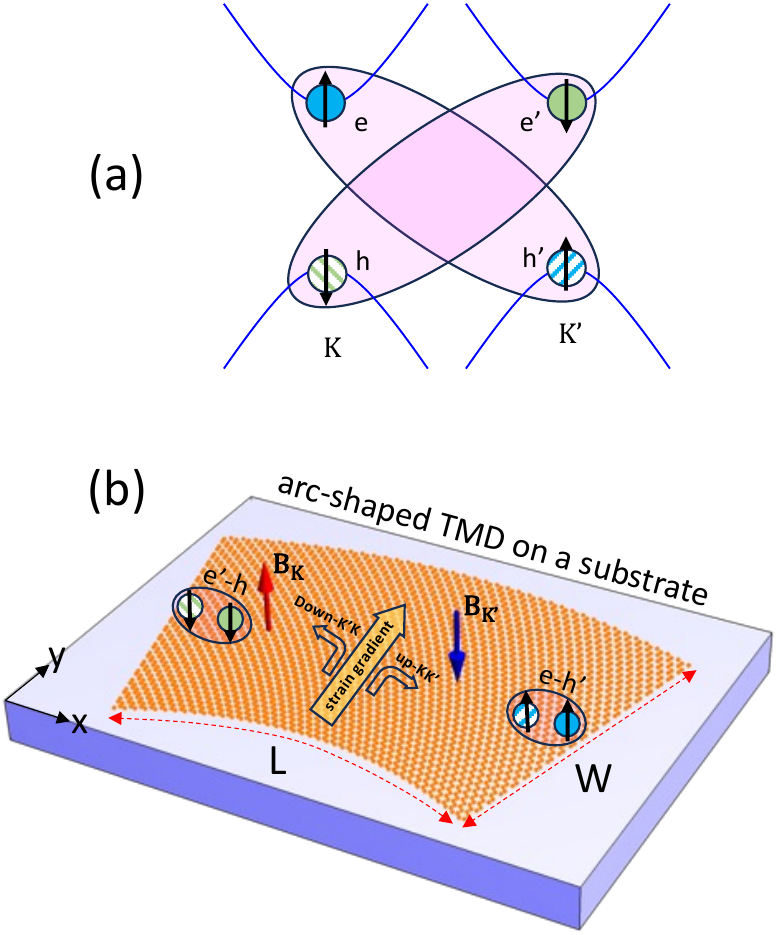}
    \caption{ (a) The scheme of the low-energy band structure with two intervalley dark excitonic states in WSe$_2$ monolayer. (b) The geometry of the proposed arc-shaped device designed to induce the exciton-mediated spin Hall effect in WSe$_2$. Introduction of a non-uniform strain gives rise to a pseudomagnetic field, generating the Lorentz forces leading to opposite transverse center-of-mass movement of up-KK$'$ and down-K$'$K-excitons. The maximum strain in the system is given by $W/2R$, where $W$ and $R$ represent the width and radius of the arc, respectively. The $x$-direction corresponds to a zigzag crystallographic  orientation. 
    \label{fig:sketch}}
\end{figure}

Among various two-dimensional (2D) materials, monolayers of transition metal dichalcogenides (TMDs) \cite{Mak2010} are of special interest. This is because of their remarkable compatibility with various semiconductor/dielectric platforms and their remarkable properties, such as peculiar 2D screening  leading to dramatic modification of the interaction potential between the charged carriers \cite{Wang2018} and  nontrivial spin and valley dynamics \cite{Yao2008,Yang2017,Zhang2019,Avsar2020}.  In particular, spin-valley coupling \cite{Xiao2012} in doped TMDs is anticipated to induce spin and valley Hall effects, whether arising from gating potentials or photoexcited density. However, since the trivial band gap is much larger than the spin-orbit coupling, MoS$_2$ and WSe$_2$ do not support quantum spin Hall effect in the insulating regime. Unlike 2H-TMDs, the 1T$'$ structure \cite{qian_Sc2014} of TMD materials is of greater topological interest as it supports the quantum spin Hall effect, even at relatively high temperatures such as $100$K in 1T$'$-WTe$_2$ \cite{Sanfeng2018,shi_SA2019,Garcia_prl_2020}.

Besides intriguing single electron properties, TMD monolayers possess a remarkable excitonic response as well. Direct energy band gap and relatively large reduced mass of an electron-hole pair combined with truly 2D nature of interacting electrons and holes results in extremely large excitonic binding energies ($\sim300$~meV) which make TMD excitons stable even at room temperatures \cite{Mak2013,Ross2013,Ugeda2014}. 

It should be noted, that the spin-orbit coupling in WSe$_2$ differs from that in MoS$_2$, which results in a reversed order of spin-polarized bands within the conduction band, so that the energy of dark intravalley spin-singlet excitons is below the energy of bright excitons \cite{Wang2018}. Moreover, dark intervalley excitons formed in WSe$_2$ exhibit even stronger binding stability (it lies 16 meV below the spin-forbidden dark exciton state in WSe$_2$) and even longer lifetime since corresponding recombination process requires an intervalley transition mediated by a valley phonon \cite{Li2019}. The intervalley excitons bear a $(+1)$ spin for electron-hole pairs originating from the K-valley electron (e) and the K$'$-valley hole (h$'$) (denoted as {\em up-KK$'$-exciton} in the further discussion), and $(-1)$ spin for electron-hole pairs arising from the K$'$-valley electron (e$'$) and the K-valley hole (h) (denoted as {\em down-K$'$K-exciton} in the further discussion). 
The high stability, long lifetime, electrical neutrality, and finite spin angular momentum of intervalley excitons in WSe$_2$ make them optimal candidates to mediate the pure spin Hall effect in 2D.

Being electrically neutral particles, excitons do not reveal any significant response to external electric and magnetic fields. However, one can use strain effects instead. 
2D materials including graphene and TMD monolayers possess very rich strain physics primarily due to the sharing three-fold symmetry point groups. Static strain leads to the appearance of a pseudo-gauge field ${\bf A}$ \cite{Vozmediano2010} in addition to the conventional scalar deformation potential $V$. Because of the time-reversal symmetry, this  pseudo-gauge field have different sign for electrons located at the valley points K and K$'$ which are time-reversal partners, ${\bf A}_{\rm K} = -{\bf A}_{\rm K'}$. An inhomogeneous strain  leads to a non-uniform pseudo-gauge field that results in a pseudomagentic field ${\bf B}_{\rm K} = -{\bf B}_{\rm K'} = {\bm \nabla}\times{\bf A}_{\rm K}$.

The case of a uniform pseudomagnetic field is particularly interesting as it allows the exploration of strain-induced cyclotron dynamics, Landau level formation and the quantum valley Hall effect \cite{Guinea2010}. Examples of experimental geometries generating such uniform pseudomagnetic fields, exceeding 100s of Tesla \cite{Levy2010,Georgi2017,Nigge2019}, include trigonal and hexagonal symmetric nano-bubbles and arc-shaped straining \cite{Low2010,Rostami2013,rostami_prb_2015}. The latter case, reported in a recent twisted bilayer graphene experiment \cite{Kapfer2023}, is particularly important, as it allows creation of a uniform pseudomagnetic fields across larger samples, which is not possible with nano-bubbles. 

In this study, we investigate the effects of the combination of strain-induced electric and magnetic fields on inter-valley exciton transport in WSe$_2$, featuring an inverted spin-split conduction band. 
Our results reveal a distinctive pure spin Hall effect for inter-valley excitons, which bear no electric or valley charge. We propose an arc-shaped WSe$_2$ setup depicted in Fig. \ref{fig:sketch}b as a suitable geometry for the observation of the effect. 
In this situation, up-KK$'$-excitons and down-K$'$K-excitons will first drift along the y axis under a gradient of the scalar potential ${\bm \nabla} V$ generated by strain, and gain a drift velocity ${\bf v}_d = - \mu_{\rm X} {\bm \nabla} V$ with $\mu_{\rm X}$ being the exciton mobility. These moving excitons are affected by the Lorentz force produced by strain-induced pseudomagnetic field and generating Hall-like transverse current.
In the stationary regime, an electron-hole pair corresponding to a up-KK$'$-exciton feels the Lorentz force 
\begin{align}
{\bf f}_{\rm up-KK'} = {\bf v}_d \times (q_{e} {\bf B}_{\rm K} + q_{h'} {\bf B}_{\rm K'})= 2 e \mu_{\rm X}  {\bm \nabla} V \times  {\bf B},
\end{align}
where we took $q_{e}=-q_{h'}=-e$ and used that ${\bf B}_{\rm K}=-{\bf B}_{\rm K'} = {\bf B}$. The Lorentz force acting at a down-K$'$K-exciton has opposite direction,  ${\bf f}_{\rm down-K'K} = - {\bf f}_{\rm up-KK'}$ and therefore up and down spins accumulate at opposite edges of the device as shown in  Fig. \ref{fig:sketch}b.

While the dynamics of pseudo-magnetoexcitons have been previously explored in strained graphene \cite{Berman2022}, here we present a robust argument supported by microscopic transport theory for the exciton-assisted spin Hall effect driven by non-uniform strain in single-layer TMD.
Note, that the proposed mechanism is distinct from conventional exciton Hall effect in TMD monolayers, which is of valley-selective nature \cite{Onga2017,Li2015,Yu2016,Huang2019,Glazov2020}.
The Lorentz force acting on an intra-valley exciton vanishes as both an electron and a hole feel the same pseudomagnetic field, and therefore the net force is zero. Note, however, that in this case pseudomagnetic field will result in the appearance of a dipole moment of a moving exciton, which leads to the appearance of the skew scattering term if an exciton interacts with an impurity and onset of the anomalous exciton Hall effect \cite{Kozin2021}.

{\em The quantitative model.}
In semiclassical picture, the dynamics of an intervalley exciton is described by the Newton's law (see Supplemental Material \cite{SM} for the detailed derivation):
\begin{align}
    \label{eq:Newton}
     \frac{ {\rm d} {\bf p}}{ {\rm d} t} = -{\bm \nabla} V ({\bf r}) -   2e\xi {\bf v}_{\bf p} \times {\bf B} ({\bf r}),
\end{align}
where the index $\xi=+$ and $\xi=-$ denotes up-KK$'$-excitons and down-K$'$K-excitons, respectively, ${\bf v}_{\bf p} =  {\bf p} / M$ is the particle velocity with $M$ being the exciton mass.
The transport properties of an ensemble of excitons are described by the Boltzmann equation for exciton distribution function $f^{\xi}_{\bf p}( {\bf r},t)$  which reads:
\begin{align}\label{eq:boltzmann}
   & \frac{\partial f^\xi_{\bf p}({\bf r},t)}{\partial t}    +\big(- {\bm \nabla} V ({\bf r}) - 2e\xi [{\bf v}_{\bf p} \times {\bf B}({\bf r})] \big) \cdot {\bm \nabla}_{\bf p} f^{\xi}_{\bf p}({\bf r},t) 
   \nonumber\\
   &+ {\bf v}_{\bf p} \cdot {\bm \nabla} f^{\xi}_{\bf p}({\bf r},t)  = -\frac{f^{\xi}_{\bf p}({\bf r},t)}{\tau_{\rm X}} -\frac{f^{\xi}_{\bf p}({\bf r},t)-\bar{f}^{\xi}_{\bf p} ( {\bf r},t)}{\tau_{\rm C}} \nonumber \\
   &  - \frac{f^{\xi}_{\bf p}({\bf r},t) 
    - f^{-\xi}_{\bf p}({\bf r},t) }{2\tau_{\rm S}}.
\end{align}
Note that $\bar{f}^\xi_{\bf p}({\bf r},t)$ is the local equilibrium density that follows Maxwell-Boltzmann distribution function. 
We used the relaxation time approximation, which accounts for exciton recombination with characteristic time $\tau_{\rm X}$ and exciton scattering on phonons and impurities with characteristic time $\tau_{\rm C}$.
The last term corresponds to exciton spin relaxation \cite{maialle1993exciton,glazov2014exciton,zhu2014exciton,jiang2021real,an2023strain}. 
The exciton density is obtained by integrating the distribution function by momentum $\textbf{p}$,
\begin{align}\label{eq:density0}
    n^\xi({\bf r},t) =  \int \frac{ {\rm d}^2 {\bf p}}{(2\pi\hbar)^2} 
    f^\xi_{\bf p}({\bf r},t) .
\end{align}
The dynamics of total exciton density $n= n^+ + n^-$ and spin density $S_z = n^+ - n^-$, related to spin polarization  as $P_z = S_z / n$ are described by the following equations (see the derivation in \cite{SM}, and the Ref. \cite{Jungel2009} therein):
\begin{align}
    \label{eq:n,Sz}
    \frac{\partial n({\bf r},t)}{\partial t} +{\bm \nabla}\cdot {\bf j}({\bf r},t)   &= -\frac{n({\bf r},t)}{\tau_{\rm X}}, \\
    \frac{\partial S_z({\bf r},t)}{\partial t} +{\bm \nabla}\cdot {\bf j}^{\rm S}({\bf r},t)   &= -\frac{S_z({\bf r},t)}{\widetilde\tau_{\rm X} }   ,
\end{align}
where $1/\widetilde{\tau}_{\rm X}=1/\tau_{\rm X}+1/\tau_{\rm S}$ and the exciton number current density reads 
\begin{align}
    \label{eq:j}
    {\bf j}({\bf r},t)  =
        \frac{\mathbfcal{J}\{n({\bf r},t) \} }{1+[\tau\omega_{\rm c}({\bf r})]^2 }     
    + \frac{\tau \omega_{\rm c}({\bf r})}{1+[\tau\omega_{\rm c}({\bf r})]^2 } \hat {\bf B}\times \mathbfcal{J}\{S_z({\bf r},t) \}, 
\end{align}
and the exciton spin current density follows 
\begin{align}\label{eq:js}
    {\bf j}^{\rm S} ({\bf r},t) =
    \frac{\mathbfcal{J}\{S_z({\bf r},t) \} }{1+[\tau\omega_{\rm c}({\bf r})]^2}   
    + \frac{\tau \omega_{\rm c}({\bf r})}{1+[\tau\omega_{\rm c}({\bf r})]^2 } \hat {\bf B}\times \mathbfcal{J}\{n({\bf r},t) \}.
\end{align}
Here, $\omega_{\rm c}({\bf r}) = 2eB({\bf r})/M$ is the exciton cyclotron frequency, with $B$ being the modulus of the strain-induced pseudomagnetic field, and $\hat{\mathbf{B}} = \mathbf{B}/B$.
The drift-diffusion current functional in the presence of a strain gradient force, ${\bm \nabla} V$, is defined as 
\begin{equation}
\mathbfcal{J}\{Y({\bf r},t)\} = -D_{\rm X} {\bm \nabla} Y({\bf r},t) - \mu_{\rm X} Y({\bf r},t) {\bm \nabla} V({\bf r}),
\end{equation}
where $\mu_{\rm X} = \tau/M$ is the exciton mobility, with 
$\tau = \left[1/\tau_{\rm C} + 1/\tau_{\rm X} + 1/(2\tau_{\rm S}) \right]^{-1} $
being an effective relaxation time. 
The diffusion constant is given by $D_{\rm X} = \mu_{\rm X} (k_{\rm B} T)$, where $k_{\rm B}$ is the Boltzmann constant and $T$ is the temperature.
It is worth to remind that considered semiclassical formalism is valid when $\omega_{\rm c} \tau \ll 1$, yielding in
$\{1+ (\tau\omega_c)^2 \}^{-1} \approx 1- (\tau\omega_c)^2 $. In the opposite regime the system is driven into a quantum regime where Landau levels are formed and a full quantum description of magneto-excitons is required \cite{Bisti2015,Berman2022}.

\begin{figure}
    \centering
    \includegraphics[width=1\linewidth]{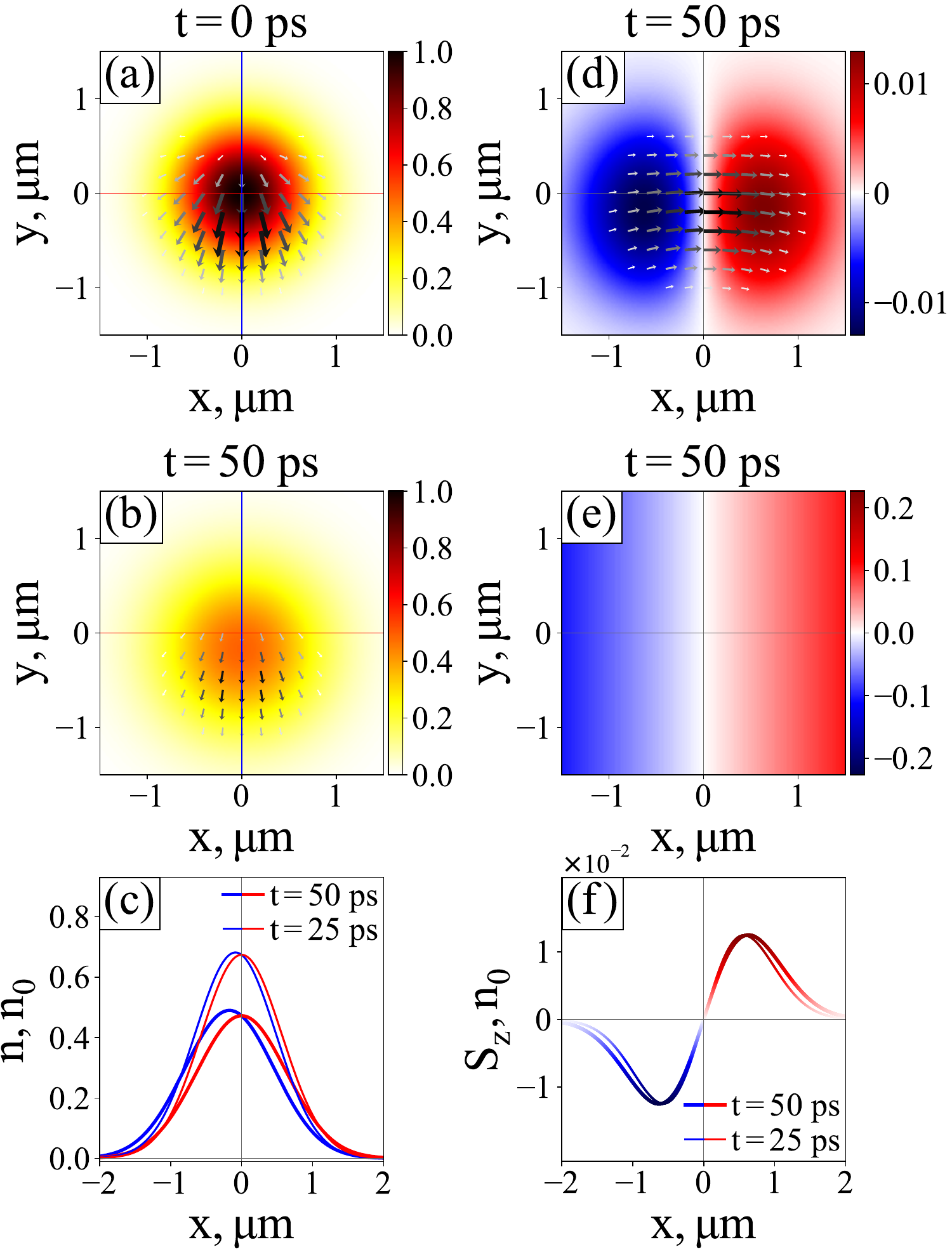}
    \caption{ Strain-induced evolution of intervalley excitons' number and spin densities. 
    Panels (a) and (b) depict snapshots of the total exciton density, denoted as $n$ (in units of $n_0$). These snapshots are taken for a system with a radius of $R = 5$ $\mu$m and a temperature of $T = 300$ K at two time points: $t = 0$ and $t = 50$ ps, respectively. The vector field arrows indicate the exciton current density, denoted as ${\bf j}$. 
    (c) Cross-sections of the total exciton density $n$ along the x-axis (depicted by the red curve) and the y-axis (depicted by the blue curve). The temporal evolution results in the broadening of the Gaussian exciton packet and its drift along the y-axis, attributed to the strain gradient ${\bm\nabla} V$. 
    Panels (d) and (e) show snapshots of the exciton spin density $S_z$ (in units of $n_0$) and the spin polarization $P_z=S_z/n$ at $t = 50$ ps, respectively. The vector field arrows in panel (d) illustrate the respective current density ${\bf j}^{\rm S}$. A valley-dependent pseudomagnetic field leads to the separation of intervalley excitons with opposite spins along the x-axis. 
    (f) Cross-section of the exciton spin density $S_z$ along the x-axis. In panels (c) and (f), the thin curves correspond to $t = 25$ ps, and the thick curves represent $t = 50$ ps.}
    \label{fig:density}
\end{figure}

In addition to the quantitative correction to the longitudinal drift-diffusion current components described by first terms of Eq. (\ref{eq:j}) and Eq. (\ref{eq:js}), pseudomagnetic field generates transverse Hall current components for both number and spin channels, as described by the second terms in Eq. (\ref{eq:j}) and Eq. (\ref{eq:js}). 

We further consider the specific arc-shaped deformation that generates a uniform pseudomagentic field. Arc-shaped strain is defined by an in-plane displacement field $(u_{x},u_{y}) = (xy/R,-x^2/2R)$, where $(x,y)$ are measured from the ribbon center and $R$ is the bending radius of the arc-shaped TMD flake. 
The scalar and gauge fields are given in terms of the strain tensor elements $u_{ij} =(\partial_i u_j +\partial_j u_i )/2 $ as $V=V_0 (u_{xx}+u_{yy})$ and ${\bf A}_{\rm K} = (\beta\phi_0/2\pi a_0) (u_{xx}-u_{yy},-2u_{xy})$ and read: 
\begin{align}\label{eq:A_V}
    {\bf A}_{\rm K} =\frac{\beta\phi_0}{2\pi a_0} \frac{y}{R} \hat {\bf x}, 
    ~~~
    V = V_0 \frac{y}{R}. 
\end{align}
where unity vector $\hat{\bf x}$ indicates a zigzag orientation in the hexagonal lattice. 
The constant pseudomagnetic field with opposite signs in the two valleys, is thus given by 
\begin{equation}
|{\bf B}| = \beta  \phi_0 /(2\pi a_0 R),
\end{equation}
 with a corresponding magnetic length of $\ell_B=\sqrt{2a_0R/\beta}$, where $\phi_0=h/2e$ is the magnetic flux quantum, $\beta\sim 3$ is the Gr\"unisen parameter describing electron-phonon coupling \cite{Rostami2018,Shahnazaryan2021}, $a=a_0\sqrt{3}$ is the lattice constant. 
For the constant pseudomagnetic field $\omega_{\rm c} ({\bf r}) = \omega_{\rm c}$ one has
${\bm \nabla} ( \tau \omega_{\rm c} D_{\rm X} {\bm\nabla} n^\xi \times \hat {\bf B} ) = 0$, so that the respective terms can be discarded in Eq. (\ref{eq:j}) and Eq. (\ref{eq:js}).

{\em Results and discussion.} 
We simulate the macroscopic transport equations \eqref{eq:n,Sz} with initial Gaussian distributions $n^\xi = n_0 e^{-(x^2+y^2)/(2\Delta^2)}$, where $\Delta$ is the characteristic size of an initial excitonic packet.
In our numerical calculations we set $\Delta = 0.5$ $\mu$m, $R = 5$ $\mu$m, $\tau_{\rm C} =0.26$ ps \cite{Cadiz2018}, $T=300$ K, $V_0 =300$ meV \cite{Moon2020}, $\tau_{\rm X} =200$ ps \cite{Li2019}, $M=0.75$ $m_0$, with $m_0$ being the free electron mass \cite{Kormanyos2015}. 
The spin relaxation time is estimated as $\tau_{\rm S} =47$ ps  \cite{SM}.
The respective pseudomagnetic field is then $B \approx 2.15$ T.

The evolution of intervalley excitons' number and spin densities under the influence of strain-induced driving fields are presented in Fig.~\ref{fig:density}. As seen in panels (a) and (b), the total density $n$ gradually broadens by diffusion, and the distribution maximum shifts in the $y$-direction due to strain-induced exciton drift. The cross-sections of the total exciton density at different time snapshots  shown in Fig.~\ref{fig:density} (c) indicate the overall longitudinal drift in the $y$-direction. 
The transverse separation of two species of intervalley excitons, characterized by exciton spin density $S_z$ and polarization $P_z$, is depicted in Fig.~\ref{fig:density} (d) and (e), respectively.
The cross-sections of the spin density in Fig.~\ref{fig:density} (f) further confirm the spin Hall current in the $x$-direction and the spin accumulation on two sides of the 2D materials almost identical to the electronic spin Hall effect \cite{Kato2004}.
Lower temperatures enhance the transverse separation of exciton spin density $S_z$ (Fig. S1 in \cite{SM}). This results from reduced diffusion and extended spin relaxation time  at low temperatures, enhancing the Hall drift. 

In order to quantify the average strength of exciton spin Hall effect relative to average longitudinal exciton number current, we introduce a Hall angle $\theta$. 
For a uniform pseudomagnetic field it reads 
\begin{equation}
    \label{eq:Hall_an}
    \tan\theta= \frac{\langle{\rm j}^{\rm S}_x\rangle}{\langle {\rm j}_y \rangle}=\frac{\int {\rm j}^{\rm S}_x {\rm d} x {\rm d} y }
    {\int {\rm j}_y {\rm d} x {\rm d} y} 
    = - \tau\omega_{\rm c} = -\frac{2e  B \tau}{M}, 
\end{equation}
With $\tan \theta = \langle{\rm j}^{\rm S}_x\rangle / \langle {\rm j}_y \rangle \propto \tau \approx \tau_{\rm C}$, where $\tau_{\rm C} \ll \tau_{\rm S} < \tau_{\rm X}$ \cite{SM}, the Hall angle $\tan \theta$ and spin polarization $S_z$ exhibit nearly linear growth with $\tau_{\rm C}$. The spatial distribution of $S_z$ across different $\tau_{\rm C}$ values is depicted in Fig. S2 \cite{SM}, highlighting the observable effect under varying levels of disorder and imperfection parameterized by $\tau_{\rm C}$. The increase in $\tau_{\rm C}$ leads to a reduced spin relaxation rate, $\tau_{\rm S} \propto 1 / \tau_{\rm C}$, following the D'yakonov-Perel' spin-relaxation mechanism \cite{Boross2013,Wu_Physics_Reports_2010,maialle1993exciton,glazov2014exciton,zhu2014exciton,jiang2021real,an2023strain}, resulting in a saturating dependence of $S_z$ on $\tau_{\rm C}$ (Fig. S2(d) in \cite{SM}).

Furthermore, as seen in Eq. \eqref{eq:Hall_an}, the Hall angle scales with the arc radius as $\tan\theta \propto B\propto 1/R$ and it can be enhanced by reducing the arc radius that implies increasing the curvature.  
Recently, arc-shaped deformation was experimentally achieved in graphene using a nano-manipulator and an atomic force microscope tip \cite{Kapfer2023}. Given a local deflection $d(x)$ along the $y$-direction in a ribbon oriented along the $x$-direction, the \emph{local} curvature radius is determined by $ R(x) = \big|{\left(1 + [\partial_x d(x)]^2\right)^{3/2}}/{\partial^2_x d(x)}\big|$.
The strain $\varepsilon(x, y) = y/R(x)$, governing both scalar and vector potentials (Eq. \eqref{eq:A_V}), reaches its maximum, $\varepsilon_{\rm max} = \pm W/2R$, at the ribbon edges $y = \pm W/2$. For instance, with a ribbon width $W = 1{\rm \mu}$ and $R = 10 {\rm \mu}$, this yields $\varepsilon_{\rm max} \approx 5\%$. The dependence of $S_z$ on the arc radius $R$ is illustrated in Fig. S3 in \cite{SM}.
Single-layer WSe$_2$ exhibits robust mechanical properties \cite{Ding_2019, Gomez_advmat_2012, Jiang_InfoMat_2020, androulidakis2018tailoring} with a reasonably large Young's Modulus ($\sim 250$ GPa \cite{Falin_acsnano_2021}), bending rigidity ($\sim 12$ eV \cite{Zhao2015}), a maximum strain tolerance of $\sim7.3\%$ in multilayer systems \cite{Zhang_APL_2016}, and a  predicted value $\sim 19.7\%$ in single-layer WSe$_2$ \cite{Falin_acsnano_2021},
securing thus the robustness for strain engineering and stability against out-of-plane buckling under arc-shaped deformations.

{\em Experimental realization.}
An intervalley exciton is the lowest energy exciton state, lying 16 meV below the spin-forbidden dark exciton state in WSe$_2$ \cite{Li2019}. 
The respective peak is clearly visible in photoluminescence spectrum at low free carrier concentration and under non-resonant optical excitation.
The optical excitation/recombination of an intervalley exciton is accompanied by the respective absorption/emission of a chiral phonon \cite{Zhu2018}, mediating the electron intervalley transition. In this process, the polarization of the absorbed/emitted photon is determined by the spin orientation of a hole, for which spin-valley locking holds. As intervalley excitons have sufficiently large lifetime of about 200 ps, in the considered configuration of arc-shaped WSe$_2$ monolayer after sufficiently long time away from the initial localized excitation pump spot the luminescence will have opposite circular polarizations, corresponding to the opposite spins of the excitons for positive and negative values of $x$-coordinate. This can be probed experimentally by near field measurement.

Moreover, spatial separation of spin polarizations generates an effective magnetic field. Indeed, the quantity $S_z(x, y) = n^{+}(x, y) - n^{-}(x, y)$ is associated with the spin angular momentum density. 
Taking into account the intervalley exciton g-factor $g\approx -12$ \cite{Li2019}, one can derive the magnetic moment density in a 2D system as $m(x, y) = g\mu_B S_z(x, y)$ where $\mu_B=57.9\times 10^{-6}$ eV/T is the Bohr magneton \cite{Liu2019}. Accounting for finite thickness of our 2D material,  $d \approx 0.7$ nm, we introduce an effective 3D magnetization, 
\begin{align}
    {\cal M}(x,y) =  \frac{g\mu_B}{d} n(x,y) P_z(x,y) \approx \frac{g\mu_B}{d} n_0 P_z(x,y) . 
\end{align}
which allows to estimate the corresponding magnetic field $B_0=\mu_0 {\cal M}$ where $\mu_0$ is the magnetic permeability of the vacuum. For realistic exciton concentrations $n_0 \sim 10^{12}$ cm$^{-2}$ this gives an estimated value of $B_0 \sim 0.002~\text{T}$, which is quite significant and falls within the range of magnetic fields typically measured in the spin Hall effect in 2D electron gas systems \cite{Kato2004}. 

The exciton spin Hall effect results in a local magnetization (and magnetic field) of sufficient magnitude to be probed by magneto-optical techniques like Faraday and Kerr rotation spectroscopy. The linear polarization rotation angle can be calculated using
\begin{align}
    \varphi_{\rm F}= {\cal V} d B_0. 
\end{align}
The Verdet constant in TMD monolayers is shown to be strongly frequency dependent and it can be as large as 
${\cal V} d \sim 5 \times 10^{-6} ~ {\rm rad/T}$ \cite{Have_prb_2019,Ferreira_prb_2011}, which yields $\varphi_{\rm F}\sim   10^{-8} ~ {\rm rad}$. Rotation angle can be further amplified by placing a monolayer in an optical cavity \cite{Kavokin1997,Da_advmat_2018}. 
Furthermore, we can enhance the effect by increasing the level of straining. The maximum strain in an arc-shaped geometry occurs at the outer edges, where the maximum strain is given by $\varepsilon_{\text{max}} = \pm W/2R$, as shown in Fig. \ref{fig:sketch}. 
When considering a fixed maximum strain, we can achieve an enhanced pseudomagnetic field by using smaller arc radii in nanoribbon systems with a narrow width $W$.  In addition, we can utilize nano-bubble structures made of WS$_2$ and WSe$_2$, which can generate substantial pseudomagnetic fields in the range of hundreds of Tesla \cite{Levy2010,Georgi2017,Nigge2019}, thereby leading to a significantly large spin accumulation due to the strain-induced exciton spin hall effect.

{\em In conclusion,} we proposed a mechanism of spin Hall effect for intervalley excitons in arc-shaped monolayers of WSe$_2$. 
In the considered geometry, the strain gradient results in both the drifting force and valley-dependent net Lorentz force acting on intervalley excitons, resulting in the onset of the transverse spin current. 
The proposed effect can be directly probed via spatially resolved near field photoluminescence spectroscopy, and pump-probe Faraday or Kerr rotation techniques. 
This effect is quite general and can also manifest itself in other indirect excitonic systems with hexagonal symmetry, such as bilayer and bulk TMD systems, where electrons and holes belong to different valley points that are time-reversal counterparts of each other. Exploring the impact of sound-induced pseudo-gauge fields, such as the acoustogalvanic effect \cite{Sukhachov2020,Bhalla2022}-recently observed in graphene \cite{Zhao2022}-on exciton spin transport, opens the possibility of extending this effect to acousto-spin modulation effects. 

The research is supported by the Ministry of Science and Higher Education of the Russian Federation (Goszadaniye) project No FSMG-2023-0011.
The derivation of theoretical model  was funded by Russian Science Foundation, project 21-72-10100.
The reported study was funded by RFBR and SC RA, project number 20-52-05005.
VS acknowledges the support of ‘Basis’ Foundation (Project No. 22-1-3-43-1).
HR acknowledges the support of Swedish Research Council (VR Starting Grant No. 2018-04252).

\clearpage

\onecolumngrid

\renewcommand{\theequation}{S\arabic{equation}}
\renewcommand{\thefigure}{S\arabic{figure}}
\setcounter{equation}{0}
\setcounter{figure}{0}

\section{Supplemental material for: \\ Exciton Spin Hall Effect In Arc-Shaped Strained WSe$_2$}

\subsection{Strain-induced force acting on exciton center-of-mass dynamics}

The classical Hamilton function for electron and hole belonging to different valleys reads ${\cal H}={\cal H}_e+{\cal H}_h + {\cal H}_{eh}$ where 
\begin{align}
    &{\cal H}_e = \frac{1}{2m_e} \left( {\bf p}_e - (-e) (+\xi) {\bf A} ({\bf r}_e) \right)^2 + E_c({\bf r}_e),
    \\
   &{\cal H}_h = \frac{1}{2m_h} \left( {\bf p}_h - (+e) (-\xi) {\bf A} ({\bf r}_h) \right)^2 - E_v({\bf r}_h),
   \\
   &{\cal H}_{eh} = {\cal H}_C({ {\bf r}_e - {\bf r}_h}).
\end{align}
Note that $\xi = + (-)$ indicates the K (K$'$) valley point, and $m_{e,h}$, ${\bf p}_{e,h}$, and ${\bf r}_{e,h}$ represent the effective mass, momentum, and coordinate for electrons and holes, respectively.
$E_c$, $E_v$ are the position dependent conduction and valence band-edge energies, ${\cal H}_C$ is the Coulomb interaction potential, ${\bf A} ({\bf r}_{e,h})$ is the strain-induced pseudogauge field.
We change the variable in terms of the center of mass and relative position vectors: 
\begin{align}
  {\bf r} = \frac{m_e{\bf r}_e + m_h {\bf r}_h}{M},~~~
  {\bm \rho} = {\bf r}_e-{\bf r}_h. 
\end{align}
Considering ${\bf p}_{e,h} = m_{e,h}\dot{\bf r}_{e,h}$, ${\bf p} =M \dot{\bf r}$ with $M=m_e+m_h$ and  ${\bf q} = \mu \dot{\bm \rho}$ with reduced mass $\mu=m_e m_h/M$ we find 
\begin{align}
{\bf p} = {\bf p}_e + {\bf p}_h,~~~
 {\bf q} = \frac{m_h {\bf p}_e - m_e {\bf p}_h}{M}. 
\end{align}
Here we are only interested in the dynamics of center of mass momentum $\bf p$.  
Using Hamilton equation of motion we can obtain the time derivative of the center of mass momentum in terms of Poisson brackets: 
\begin{align}
    \frac{ {\rm d} {\bf p}}{ {\rm d} t} = \{{\bf p},{\cal H}\} = \{{\bf p}_e,{\cal H}_e\} +\{{\bf p}_h,{\cal H}_h\}  + \{{\bf p},{\cal H}_{eh}\}.  
\end{align}
Note that $\{{\bf p}_h,{\cal H}_e\}=\{{\bf p}_e,{\cal H}_h\}=0$ since electron and hole are independent particles.  
\begin{align}
    &\{{\bf p}_e,{\cal H}_e\} =   -e\xi \dot{\bf r}_e \times {\bf B}({\bf r}_e)-\partial_{{\bf r}_e} E_c({\bf r}_e), 
     \\
    &  \{{\bf p}_h,{\cal H}_h\} =  -e\xi \dot{\bf r}_h \times {\bf B}({\bf r}_h) +  \partial_{{\bf r}_h} E_v({\bf r}_h),
    \\
    & \{{\bf p},{\cal H}_{eh}\} = -\partial_{\bf r} {\cal H}_C ({\bm \rho})=0.
\end{align}
Note that ${\bf B}({\bf r}_{e,h}) = \partial_{{\bf r}_{e,h}}\times {\bf A}({\bf r}_{e,h})$. The last term in the above relation implies that strain-induced center of mass force is independent from electron-hole interaction potential and therefore the center of mass force is give by 
\begin{align}
    \frac{ {\rm d} {\bf p}}{ {\rm d} t} =  -e\xi[\dot{\bf r}_e \times {\bf B}({\bf r}_e)+ \dot{\bf r}_h \times {\bf B}({\bf r}_h)] -\partial_{{\bf r}_e} E_c({\bf r}_e) +\partial_{{\bf r}_h} E_v({\bf r}_h).
\end{align}
We can rewrite this force in the center of mass frame: 
\begin{align}
   & {\bf r}_e= {\bf r}+\frac{m_h}{M} {\bm \rho}, ~~~ \partial_{{\bf r}_e} = \partial_{\bf r}+\frac{m_h}{M} \partial_{\bm \rho},
    \\
   & {\bf r}_h= {\bf r}-\frac{m_e}{M} {\bm \rho}, ~~~\partial_{{\bf r}_h}= \partial_{\bf r}-\frac{m_e}{M} \partial_{\bm\rho}. 
\end{align}
Therefore, the effective Lorenz force reads 
\begin{align}
&\dot{\bf r}_e \times {\bf B}({\bf r}_e)+ \dot{\bf r}_h \times {\bf B}({\bf r}_h)
= \dot{\bf r} \times \left [ {\bf B}\left({\bf r}+\frac{m_h}{M} {\bm \rho} \right)+ {\bf B}\left({\bf r}-\frac{m_e}{M} {\bm\rho} \right) \right] + 
 \dot{\bm \rho} \times \frac{ m_h {\bf B}\left({\bf r}+\frac{m_h}{M} {\bm\rho} \right) - m_e {\bf B}\left({\bf r}-\frac{m_e}{M} {\bm \rho} \right) }{M}.
\end{align}
Now, we make an approximation and neglect the dependence of the magnetic field on the relative distance ${\bm\rho}$. This is a valid approximation since the wavelength of deformation is usually much larger than the size of excitons. In particular, for a uniform magnetic field, this is an exact treatment. Therefore, we obtain 

\begin{align}
&\dot{\bf r}_e \times {\bf B}({\bf r}_e)+ \dot{\bf r}_h \times {\bf B}({\bf r}_h)
\approx 2 \dot{\bf r} \times {\bf B}({\bf r}) + \frac{ m_h-m_e  }{M}
 \dot{\bm\rho} \times  {\bf B}({\bf r}).
\end{align}
Similarly, the scalar potential force is given as follows
\begin{align}
    &\partial_{{\bf r}_e} E_c({\bf r}_e) = \partial_{\bf r} E_c\left({\bf r}+\frac{m_h}{M} {\bm \rho}\right)+\frac{m_h}{M} \partial_{\bm \rho} E_c\left({\bf r}+\frac{m_h}{M} {\bm \rho}\right),
    \\
    &\partial_{{\bf r}_h} E_v({\bf r}_h) = \partial_{\bf r} E_v\left({\bf r}+\frac{m_e}{M} {\bm \rho}\right)-\frac{m_e}{M} \partial_{\bm \rho} E_v\left({\bf r}-\frac{m_e}{M} {\bm \rho}\right).
\end{align}
Within the same approximation of long wavelength deformation, we obtain
\begin{align}
    &\partial_{{\bf r}_e} E_c({\bf r}_e) \approx  \partial_{\bf r} E_c({\bf r}),
    \\
    &\partial_{{\bf r}_h} E_v({\bf r}_h) \approx \partial_{\bf r} E_v({\bf r}).
\end{align}
Therefore, the center of mass force reads 
\begin{align}
    \frac{ {\rm d} {\bf p}}{ {\rm d} t} \approx -2e\xi \dot{\bf r} \times {\bf B}({\bf r})  
 -e\xi \frac{ m_h-m_e  }{M}
 \dot{\bm\rho} \times  {\bf B}({\bf r})  -\partial_{\bf r} [E_c({\bf r})-E_v({\bf r})].
\end{align}
We neglect the relative dynamics of electrons and holes, as the center-of-mass movement is significantly larger than the exciton size. Therefore, we can safely omit the term dependent on $\dot{\bm \rho}$.  Accordingly, we finally obtain 

\begin{align}
   \frac{ {\rm d} {\bf p}}{ {\rm d} t} \approx -2e\xi {\bf v}_{\bf p} \times {\bf B}({\bf r})   -\partial_{\bf r} V({\bf r}),
\end{align}
where ${\bf v}_{\bf p} = \dot{\bf r}$ is the center of mass velocity and $V({\bf r})=E_c({\bf r})-E_v({\bf r})$ stands for the strain-induced band-gap renormalization.


\subsection{Derivation of semiclassical transport equations}

At quasiclassical limit, the gas of KK$'$ and K$'$K intervalley excitons can be characterized by distribution function $f^{\xi}_{\bf p}( {\bf r},t)$. At {\em local} equilibrium the distribution function follows Maxwell-Boltzmann type with the local particle density (instead of the global one) as the pre-factor:  
\begin{equation}
    \bar{f}^{\xi}_{\bf p}( {\bf r},t) = \frac{n^{\xi}({\bf r},t)}{N} e^{-\frac{(p^2/2M)}{ k_{\rm B} T}},
\end{equation}
where $N = (M k_{\rm B} T) / (2\pi \hbar^2) $. 
Here $n^{\xi}({\bf r},t)$ is the intervalley exciton density, reading as
\begin{equation}\label{eq:density}
    n^{\xi}({\bf r},t) = \int \frac{d^2 {\bf p}}{(2\pi \hbar)^2} \bar{f}^{\xi}_{\bf p}({\bf r},t).
\end{equation}
Later, we will see that the correction to the distribution function has odd parity in momentum and, therefore, it does not contribute to the particle density but gives rise to the current density. In this regard, we can replace $\bar{f}^{\xi}$ with $f^{\xi}$ in Eq. \eqref{eq:density} as well as Eq. (4) of the main text. 
The Boltzmann equation (3) of the main text for exciton gas distribution reads
\begin{align}
    \frac{ {\rm d} f^{\xi}_{\bf p}({\bf r},t) }{ {\rm d} t} 
    &= \frac{\partial f^{\xi}_{\bf p}({\bf r},t) }{\partial t} 
    + {\bf v}_{\bf p} \cdot {\bm\nabla} f^{\xi}_{\bf p}({\bf r},t) 
    -  \left( {\bm\nabla} V +2 e {\bf v}_{\bf p} \times {\bf B}({\bf r}) \right) \cdot {\bm\nabla}_{\bf p} f^{\xi}_{\bf p}({\bf r},t)  
    \notag \\
   & = -\frac{f^{\xi}_{\bf p}({\bf r},t) }{\tau_{\rm X}}  -\frac{f^{\xi}_{\bf p}({\bf r},t) -\bar{f}^{\xi}_{\bf p}( {\bf r},t) }{\tau_{\rm C}} 
    - \frac{f^{\xi}_{\bf p}({\bf r},t) 
    - f^{\xi'}_{\bf p}({\bf r},t)}{2\tau_{\rm S}} ,
\end{align}
where we denote $f^{\xi'}_{\bf p}({\bf r},t) = f^{-\xi}_{\bf p}({\bf r},t)$. Given by the considerable energy gap between the intervalley and intravalley excitons, we consider only the relaxation between the lowest energy intervalley excitons. 
From now on, we omit the superscripts $\xi$ for the shorthand writing, and consider the case of KK$'$ intervalley excitons. The derivation for K$'$K exciton is identical.
For the derivation of macroscopic transport equations we proceed with first introducing dimensionless variables.
One can define the mean free path as $\lambda_{\rm  C} = u \tau_{\rm  C}$, where $u = \sqrt{k_{\rm B} T /M}$ is the thermal velocity. 
Denoting the sample length as $L$, we define the characteristic space and time scales as 
$\lambda_0 = \sqrt{L \lambda_{\rm  C}}$, $\tau_0 = L/u$. Furthermore, we introduce dimensionless variable $\alpha = \lambda_{\rm  C} / \lambda_0 = \sqrt{\lambda_{\rm  C} / L} = \sqrt{\tau_{\rm  C} / \tau_0}$, and the momentum unit reads $p_0 = M u$.
Accordingly, we rewrite the model parameters in terms of this physical units and dimensionless quantities: 
\begin{equation}
    {\bf r} = \lambda_0 \Tilde{\bf r} , \qquad
    t = \tau_0 \Tilde{t}, \qquad
    {\bf p} = p_0 \Tilde{\bf p}, \qquad
    {\bf v}_{\bf p} = \frac{\bf p}{M} = u \Tilde{\bf p}, \qquad
    V = (k_{\rm B} T) \Tilde{V} .
\end{equation}
Using this parametrization, the Boltzmann kinetic equation reads 
\begin{align}
    \frac{1}{\tau_0} \frac{\partial f}{\partial \Tilde{t} }
    + \frac{u }{\lambda_0} \Tilde{ \bf p} \cdot \partial_{\Tilde{\bf r}} f 
    -\left( \frac{k_B T}{ \lambda_0} \frac{1}{M u} \partial_{\Tilde{\bf r}} \Tilde{V} 
    + 2 e \frac{1}{M u} u [\Tilde{\bf p} \times {\bf B} (\Tilde{\bf r}) ] \right) \cdot \partial_{\Tilde{\bf p}} f
    +\frac{f}{\tau_{\rm X}}
    =-\frac{f-\bar{f}}{\tau_{\rm C}}
    -  \frac{f - f^{'}}{2\tau_{\rm S} } .
\end{align}
Multiplying by $\tau_{\rm C}$, we note that 
\begin{align}
    \frac{\tau_{\rm C}}{\tau_0} = \alpha^2,~~~
    \frac{u \tau_{\rm C}}{\lambda_0}  = \alpha,~~~ 
    \frac{\tau_{\rm C} k_B T }{\lambda_0 M u}  = \alpha. 
\end{align}
Recalling that ${\bf B} (\Tilde{\bf r}) = B (\Tilde{\bf r}) \hat {\bf B}$, and cyclotron frequency $\omega_c  (\Tilde{\bf r}) = 2e B (\Tilde{\bf r}) / M $,  we finally obtain 
\begin{align}\label{eq:kinetic-dimensionless}
    \alpha^2 \frac{\partial f}{\partial \Tilde{t} }
    + \alpha \Tilde{ \bf p} \cdot \partial_{\Tilde{\bf r}} f 
    -\left( \alpha \partial_{\Tilde{\bf r}} \Tilde{V} 
    + \tau_{\rm C} \omega_c (\Tilde{\bf r}) [\Tilde{\bf p} \times \hat {\bf B} ] \right) \cdot \partial_{\Tilde{\bf p}} f 
    +\frac{\tau_{\rm C}}{\tau_{\rm X}} f
    = -f +\bar{f} 
    - \frac{\tau_C}{2\tau_{\rm S} } (f - f^{'}).
\end{align}
In the regime of weak magnetic field  $\tau_{\rm C} \omega_c / \alpha = \sqrt{\tau_C \tau_0} \omega_c \ll 1$.
We then apply a Chapman–Enskog expansion \cite{Jungel2009} where the population can be written by the following ansatz: 
\begin{align}\label{eq:Chapman–Enskog}
  f_{\Tilde{\bf p}} (\Tilde{\bf r}, \Tilde{t}) = \Tilde{n} (\Tilde{\bf r}, \Tilde{t}) f^{\rm MB}_{\Tilde{\bf p}} e^{-\frac{\tau_0}{\tau_X}\Tilde{t}} + \alpha g_{\Tilde{\bf p}}(\Tilde{\bf r}, \Tilde{t}) e^{-\frac{\tau_0}{\tau_X}\Tilde{t}},  
\end{align}
where $f^{\rm MB}_{\Tilde{\bf p}} = 2\pi e^{-\Tilde{p}^2/2}$ is the Maxwell-Boltzmann distribution in terms of dimensionless variables. The first term represents the local equilibrium distribution, taking into account the finite recombination lifetime of excitons $\tau_{\rm X}$. Note that $n (\Tilde{\bf r}, \Tilde{t}) =\Tilde{n} (\Tilde{\bf r}, \Tilde{t}) e^{-\frac{\tau_0}{\tau_X}\Tilde{t}}$. The function $g_\alpha$ denotes the correction to the distribution function caused by external perturbations, which, in this case, are strain-induced fields. 
Plugging Chapman–Enskog ansatz of distribution function (\ref{eq:Chapman–Enskog}) into the kinetic equation (\ref{eq:kinetic-dimensionless}), we get 
\begin{align}
    \label{eq:Boltzmann-alpha}
   & \alpha \left( \frac{\partial \Tilde{n} }{\partial \Tilde{t} } 
    -\frac{\tau_0}{\tau_{\rm X}} \Tilde{n} \right)f^{\rm MB}
    +\alpha^2 \left( \frac{\partial g }{\partial \Tilde{t} }  - \frac{\tau_0}{\tau_{\rm X}} g \right)
    + \left( f^{\rm MB} \Tilde{ \bf p} \cdot \partial_{\Tilde{\bf r}} \Tilde{n} 
    -\Tilde{n} \partial_{\Tilde{\bf r}} \Tilde{V} \cdot  \partial_{\Tilde{\bf p}} f^{\rm MB} \right)
    + \alpha \left( \Tilde{ \bf p} \cdot \partial_{\Tilde{\bf r}} g
    -\partial_{\Tilde{\bf r}} \Tilde{V} \cdot \partial_{\Tilde{\bf p}} g  \right) 
    \notag \\
    &- \frac{\tau_{\rm C} \omega_c (\Tilde{\bf r})}{\alpha}  [\Tilde{\bf p} \times \hat {\bf B} ] \cdot \Tilde{n} \partial_{\Tilde{\bf p}} f^{\rm MB}
    - \tau_{\rm C} \omega_c (\Tilde{\bf r})   [\Tilde{\bf p} \times \hat {\bf B} ] \cdot \partial_{\Tilde{\bf p}} g
    +\frac{\tau_{\rm C}}{\tau_X \alpha} \Tilde{n} f^{\rm MB} +\frac{\tau_{\rm C}}{\tau_{\rm X} } g
    = -g
    -  \frac{\tau_C}{2\tau_{\rm S} \alpha} (\Tilde{n} f^{\rm MB} +\alpha g 
    - \Tilde{n}' f^{\rm MB '} -\alpha g' ).
\end{align}
We note that 
$\tau_{\rm C}/(\tau_{\rm X} \alpha) = (\alpha \tau_0)/\tau_{\rm X}$, leading to the cancellation of respective terms.
In addition,
$[\Tilde{\bf p} \times \hat {\bf B}]  \cdot   \partial_{\Tilde{\bf p}} f^{\rm MB} = - [\Tilde{\bf p} \times \hat {\bf B} ]  \cdot   \Tilde{\bf p} f^{\rm MB} =0$. 
We are looking for the leading-order correction that can be obtain considering the limit $\alpha\ll1$ \cite{Jungel2009}. Therefore, we obtain 
\begin{align}
    g-
    \tau \omega_c [\Tilde{\bf p} \times \hat {\bf B} ] \cdot \partial_{\Tilde{\bf p}} g = {\cal S}_{\Tilde{\bf p}} = -\frac{\tau}{\tau_{\rm C}} \left( f^{\rm MB} \Tilde{ \bf p} \cdot \partial_{\Tilde{\bf r}} \Tilde{n} 
    -\Tilde{n} \partial_{\Tilde{\bf r}} \Tilde{V} \cdot  \partial_{\Tilde{\bf p}} f^{\rm MB} \right)
    - \frac{\tau}{2\tau_{\rm S}\alpha} (\Tilde{n} f^{\rm MB} 
    -\Tilde{n}' f^{\rm MB '} -\alpha g').
\end{align}
where $\tau = \left[1/\tau_{\rm C} + 1/\tau_{\rm X} + 1/(2\tau_{\rm S}) \right]^{-1} $
is an effective relaxation time.
This equation is a first-order differential equation for $g$ with a source function $S_{\Tilde{\bf p}}$. We first consider the homogeneous equation by setting ${\cal S}_{\Tilde{\bf p}}=0$: 
\begin{align}
    g -  \tau \omega_c  [\Tilde{\bf p} \times \hat {\bf B} ] \cdot \partial_{\Tilde{\bf p}} g = 0 .
\end{align}
In polar coordinates $\Tilde{ \bf p} = (\Tilde{p} \cos\phi, \Tilde{p}\sin\phi)$, one can find 
\begin{align}
     g +  \tau \omega_c  \frac{\partial g}{\partial \phi} = 0 ,
\end{align}
yielding in the following general solution 
\begin{align}
  g = C_0 e^{ - \frac{\phi}{\tau \omega_c }}.   
\end{align}
where $C_0$ is a constant. 
We now use the above ansatz for $g$ where we replace $C_0$ with $C$ that is a function of position, momentum and time. 
Given that $\partial_{\Tilde{\bf p}} f^{\rm MB} = - \Tilde{\bf p} f^{\rm MB}$, we obtain the angular derivative of $C$: 
\begin{align}
    \frac{\partial C}{\partial \phi} = - \frac{e^{\frac{\phi}{\tau\omega_c}} }{\tau_C \omega_c}  
    \Tilde{ \bf p} \cdot \left(   \partial_{\Tilde{\bf r}} \Tilde{n} 
    +\Tilde{n} \partial_{\Tilde{\bf r}} \Tilde{V}   \right) f^{\rm MB}
    - \frac{e^{\frac{\phi}{\tau\omega_c}}}{2\tau_{\rm S}\omega_C\alpha} (\Tilde{n} f^{\rm MB} 
    - \Tilde{n}' f^{\rm MB '} -\alpha g').
\end{align}
Integrating the last equation we get:
\begin{align}
   C = C_0 - \frac{\tau}{\tau_{\rm C}}
   \frac{ e^{\frac{\phi}{\tau\omega_c}}}{1+ (\tau \omega_c)^2}
   \left[ \Tilde{ \bf p} \cdot \left(   \partial_{\Tilde{\bf r}} \Tilde{n} 
    +\Tilde{n} \partial_{\Tilde{\bf r}} \Tilde{V}   \right) 
    - (\tau \omega_c )
     \left([\Tilde{\bf p} \times \partial_{\Tilde{\bf r}}]_z \Tilde{n}  
    + \Tilde{n}   [\Tilde{\bf p} \times \partial_{\Tilde{\bf r}}]_z \Tilde{V} \right)
    \right] f^{\rm MB}
     - \frac{\tau e^{\frac{\phi}{\tau\omega_c}}}{2\tau_{\rm S}\alpha} (\Tilde{n} f^{\rm MB} 
    - \Tilde{n}' f^{\rm MB '} -\alpha g').
\end{align}
The constant $C_0$ contributes to the {\em general} solution in the absence of strain gradient. However, in the local equilibrium, we must have $g=0$, which implies $C_0=0$. Accordingly, we obtain the {\em particular} solution of $g$ as follows
\begin{align}
    g = - \frac{\tau}{\tau_{\rm C}} \frac{f^{\rm MB}}{1 + (\tau \omega_c)^2} 
    \left[   \Tilde{\bf p} \cdot \partial_{\Tilde{\bf r}} \Tilde{n}   
    + (\partial_{\Tilde{\bf r}} \Tilde{V}) \cdot \Tilde{\bf p} \Tilde{n}  
    -  (\tau \omega_c)  \left( [\Tilde{\bf p} \times \partial_{\Tilde{\bf r}}]_z \Tilde{n}  
    + \Tilde{n}   [\Tilde{\bf p} \times \partial_{\Tilde{\bf r}}]_z \Tilde{V} \right) \right]
     \frac{\tau }{2\tau_{\rm S}\alpha} (\Tilde{n} f^{\rm MB} 
    - \Tilde{n}' f^{\rm MB '} -\alpha g'),
\end{align}
which can be rewritten as
\begin{align}\label{eq:g}
    g = - \frac{\tau}{\tau_{\rm C}} \frac{f^{\rm MB}}{1 + (\tau \omega_c)^2} 
    \left[   \Tilde{\bf p} \cdot \partial_{\Tilde{\bf r}} \Tilde{n}   
    + (\partial_{\Tilde{\bf r}} \Tilde{V}) \cdot \Tilde{\bf p} \Tilde{n}  
    -  (\tau \omega_c)  \left( [\hat{\bf B}\times \Tilde{\bf p}] \cdot {\partial}_{\Tilde{\bf r}} \Tilde{n}  
    + \Tilde{n}   [\hat{\bf B}\times \Tilde{\bf p}] \cdot {\partial}_{\Tilde{\bf r}}\Tilde{V} \right) \right]
     - \frac{\tau }{2\tau_{\rm S}\alpha} (\Tilde{n} f^{\rm MB} 
    - \Tilde{n}' f^{\rm MB '} -\alpha g'),
\end{align}
We next perform momentum integration of Eq. \eqref{eq:Boltzmann-alpha} which leads to 
\begin{align}
   & \langle f^{\rm MB} \rangle \frac{\partial \Tilde{n}}{\partial \Tilde{t} } + \alpha \frac{\partial \langle g \rangle}{\partial \Tilde{t} }  
    + \frac{\partial_{\Tilde{\bf r}}\Tilde{n} \cdot \langle \Tilde{ \bf p} f^{\rm MB}  \rangle
    -\partial_{\Tilde{\bf r}} \Tilde{V} \Tilde{n} \cdot \langle \partial_{\Tilde{\bf p}}f^{\rm MB}  \rangle }{\alpha} 
    +   \langle \Tilde{ \bf p} \cdot \partial_{\Tilde{\bf r}} g \rangle
    -\partial_{\Tilde{\bf r}} \Tilde{V} \cdot \langle \partial_{\Tilde{\bf p}} g \rangle  
    - \frac{\tau_{\rm C} \omega_c \langle [\Tilde{\bf p} \times \hat {\bf B}]  \cdot \partial_{\Tilde{\bf p}} g \rangle}{\alpha}  
    \notag \\
    &= -\frac{\langle g \rangle}{\alpha} 
    - \frac{\tau_C}{2\tau_{\rm S}\alpha^2} (\Tilde{n} \langle f^{\rm MB} \rangle 
    - \Tilde{n}' \langle f^{\rm MB '} \rangle
    +\alpha \langle g \rangle -\alpha \langle g' \rangle).
\end{align}
Where $\langle O \rangle \equiv \sum_{\bf p} O_{\bf p}$. We note that $\langle f^{\rm MB} \rangle = 1$, and the terms $\langle \Tilde{\bf p}  f^{\rm MB} \rangle$, $\langle \partial_{\Tilde{\bf p}} f^{\rm MB} \rangle$, $\langle g \rangle$, and $\langle [\Tilde{\bf p} \times \hat{\bf B}] \cdot \partial_{\Tilde{\bf p}} g \rangle$ are odd in $\Tilde{\bf p}$ and thus vanish. Moreover, the boundary integral $\langle \partial_{\Tilde{\bf p}} g \rangle = \sum_{\bf p} \partial_{\Tilde{\bf p}} g_{\Tilde{\bf p}} = 0$.
Therefore, we obtain 
\begin{align}
     \frac{\partial \Tilde{n} }{\partial \Tilde{t} } 
    + \langle \Tilde{ \bf p} \cdot \partial_{\Tilde{\bf r}} g \rangle
    = - \frac{\tau_0}{2\tau_{\rm S}} (\Tilde{n} 
    - \Tilde{n}'  ).
\end{align}
Plugging the expression of $g$ given in Eq. \eqref{eq:g} into the above relation, we achieve 
\begin{align}
     \frac{\partial \Tilde{n} }{\partial \Tilde{t} } 
    & - \frac{\tau}{\tau_{\rm C}} \partial_{\Tilde{\bf r}} \cdot \left(
    \frac{1}{1+(\tau\omega_c)^2} 
    \left[ \left\langle \Tilde{ \bf p}  \left( \Tilde{ \bf p} \cdot \partial_{\Tilde{\bf r}} \Tilde{n} f^{\rm MB} \right) \right\rangle
    + \left\langle \Tilde{ \bf p} \left(\partial_{\Tilde{\bf r}} \Tilde{V} \cdot \Tilde{\bf p} \Tilde{n} f^{\rm MB} \right)\right \rangle \right. \right. \notag \\
    & \left. \left. - \tau \omega_c \left\langle \Tilde{ \bf p} \cdot \left( [\hat {\bf B} \times \Tilde{\bf p} ] \cdot \partial_{\Tilde{\bf r}} \Tilde{n} f^{\rm MB} \right)\right\rangle
    -\tau \omega_c  \left\langle \Tilde{ \bf p} \cdot \left( \Tilde{n} f^{\rm MB} [\hat {\bf B} \times \Tilde{\bf p}] \cdot \partial_{\Tilde{\bf r}} \Tilde{V} \right) \right\rangle
    \right] \right) \notag \\
    & - \frac{\tau }{2\tau_{\rm S}\alpha} 
    \partial_{\Tilde{\bf r}} \left( \Tilde{n} \langle \Tilde{\bf p}  f^{\rm MB}  \rangle
    - \Tilde{n}' \langle \Tilde{\bf p}  f^{\rm MB '} \rangle -\alpha \langle \Tilde{\bf p} g' \rangle \right)
    = - \frac{\tau_0}{2\tau_{\rm S}} (\Tilde{n} 
    - \Tilde{n}'  ).
\end{align}
The first two terms in the bracket result in:
\begin{align}
    \langle \Tilde{ \bf p}  ( \Tilde{ \bf p} \cdot \partial_{\Tilde{\bf r}} \Tilde{n} f^{\rm MB} ) \rangle 
    &= \partial_{\Tilde{\bf r}} \Tilde{n}, \notag \\
    \langle \Tilde{ \bf p} (\partial_{\Tilde{\bf r}} \Tilde{V} \cdot \Tilde{\bf p} \Tilde{n} f^{\rm MB}) \rangle   
    &= \Tilde{n} \partial_{\Tilde{\bf r}} \Tilde{V} .  
\end{align}
The third term is evaluated as
\begin{align}
    \tau \omega_c   \left\langle \Tilde{ \bf p} \cdot \left( [\hat {\bf B} \times \Tilde{\bf p} ] \cdot \partial_{\Tilde{\bf r}} \Tilde{n} f^{\rm MB} \right) \right\rangle
    =  \tau \omega_c  \left( \hat{\bf x} \frac{\partial \Tilde{n}}{\partial \Tilde{y}}
    -\hat{\bf y} \frac{\partial \Tilde{n}}{\partial \Tilde{x}} \right)
    =  \tau \omega_c  [\partial_{\Tilde{\bf r}} \Tilde{n} \times \hat {\bf B}] .
\end{align}
The same manner for the last term one has
\begin{align}
    \tau \omega_c \left \langle \Tilde{ \bf p} \cdot \left( [\hat {\bf B} \times \Tilde{\bf p} ] \cdot (\partial_{\Tilde{\bf r}} \Tilde{V}) \Tilde{n} \bar{f}_0 \right) \right\rangle
    = \tau \omega_c [\partial_{\Tilde{\bf r}} \Tilde{V} \times \hat {\bf B}] \Tilde{n} .
\end{align}
The terms $\langle \Tilde{\bf p}  f^{\rm MB}  \rangle$, $\langle \Tilde{\bf p}  f^{\rm MB'} \rangle$ are odd in $\Tilde{\bf p}$ and vanish.
We neglect the term $ \tau  \langle \Tilde{\bf p} g' \rangle / (2\tau_{\rm S})$,
given by $ \tau^2 / (\tau_{\rm C} \tau_{\rm S}) \ll 1$.
The rate equation for $\Tilde{n}$ then reads
\begin{align}
    \frac{\partial \Tilde{n} }{\partial \Tilde{t} } 
    = \frac{\tau}{\tau_{\rm C}}\partial_{\Tilde{\bf r}} 
    \cdot
    \left(
    \frac{1}{1+(\tau\omega_c)^2}
    \left(  \partial_{\Tilde{\bf r}} \Tilde{n} + \Tilde{n} \partial_{\Tilde{\bf r}} \Tilde{V} 
    -\tau \omega_c [\partial_{\Tilde{\bf r}} \Tilde{n} \times \hat {\bf B}] 
    -\tau \omega_c [\partial_{\Tilde{\bf r}} \Tilde{V} \times \hat {\bf B}] \Tilde{n} \right) 
    \right) 
    - \frac{\tau_0}{2\tau_{\rm S}} (\Tilde{n} 
    - \Tilde{n}'  ).
\end{align}
We recall that $\Tilde{n} = n e^{\frac{\tau_0}{\tau_X}\Tilde{t}}$, resulting in 
\begin{align}
    \frac{\partial n }{\partial \Tilde{t} }   
    + \partial_{\Tilde{\bf r}} \cdot \Tilde{ {\bf J}} 
     =- \frac{\tau_0}{\tau_{\rm X}} n 
     - \frac{\tau_0}{2\tau_{\rm S}} (n - n'  ), 
\end{align}
with the following current density   
\begin{align}
    {\Tilde{ \bf J}}  = -\frac{\tau/\tau_{\rm C}}{1+(\tau\omega_c)^2}
    \left(  \partial_{\Tilde{\bf r}} n + n \partial_{\Tilde{\bf r}} \Tilde{V} 
    -\tau \omega_c  [\partial_{\Tilde{\bf r}} n \times \hat {\bf B}] 
    -\tau \omega_c  [\partial_{\Tilde{\bf r}} \Tilde{V} \times \hat {\bf B}] n \right) .
\end{align}
Finally, by restoring the dimensional units, we obtain the density rate equation and corresponding current density as follows.
\begin{align}
    \frac{\partial n }{\partial t } 
    +\partial_{\bf r} \cdot {\bf J} 
     =- \frac{n}{\tau_{\rm X}}
    - \frac{n - n'}{2\tau_{\rm S}},
\end{align}
where the current density reads 
\begin{align}
    {\bf J} = - 
    \frac{D_{\rm X} \partial_{\bf r} n + \mu_{\rm X} n \partial_{\bf r} V }{1+ (\tau\omega_c)^2}
    +\frac{\tau \omega_c}{1+ (\tau\omega_c)^2} \left( D_{\rm X} \partial_{\bf r} n 
    + \mu_{\rm X} n  \partial_{\bf r} V  \right) \times \hat {\bf B} . 
\end{align}
Here $\mu_{\rm X} =\tau/M$,  $D_{\rm X} = \mu_{\rm X} (k_{\rm B} T)$ indicate the mobility, and diffusion coefficient of excitons, respectively. 

\subsection{Estimation of the exciton spin relaxation time}

The spin relaxation of two-dimensional excitons is well studied both for quantum wells \cite{maialle1993exciton} and atomic monolayers \cite{glazov2014exciton}. 
The relaxation time for thermalized excitons can be estimated as based on the D'yakonov-Perel' mechanism of spin-relaxation time \cite{Boross2013,Wu_Physics_Reports_2010} and it follows \cite{zhu2014exciton}:
\begin{equation}
    \tau_{\rm S} = \frac{1}{2\tau_C M k_B T}
    \left (\frac{\epsilon+1}{\sqrt{\epsilon} +1}  \frac{ \tau_X E_X }{c} \right)^2,
\end{equation}
where $\epsilon$ is the effective dielectric permittivity of surrounding media, $c$ is the speed of light, $M$ is the exciton mass, and $E_X$ is the resonance energy of intervalley exciton. 
The expression is valid at the conditions $k_B T \tau_{\rm C} / \hbar  \gg 1$, $\tau_{\rm C} \ll \tau_{\rm S} < \tau_{\rm X}$.
By setting $\epsilon = 3$ and $M = 0.75m_0$ in terms of the free electron mass ($m_0$), and using the experimentally estimated intervalley exciton energy $E_X = 1681$ meV  \cite{Li2019}, we calculate $\tau_{\rm S} \approx 47$ ps at room temperature ($T = 300$ K).

\begin{figure}[h!]
    \centering
    \includegraphics[width = \linewidth]{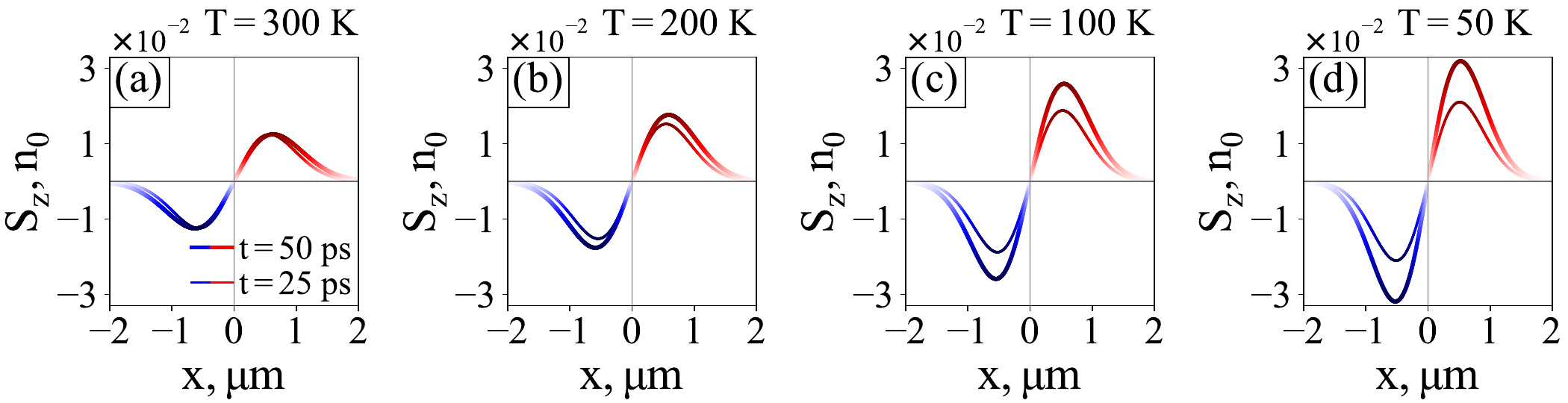}
    \caption{Spatial distribution of exciton spin density $S_z$ at different temperatures: (a) $T = 300$ K, (b) $T = 200$ K, (c) $T = 100$ K, (d) $T = 50$ K. The decrease of temperature suppresses the diffusion, and increases the spin relaxation time, both leading to the enhancement of spatial separation of exciton spin density.}
    \label{fig:S1}
\end{figure}

\begin{figure}[h!]
    \centering
    \includegraphics[width = \linewidth]{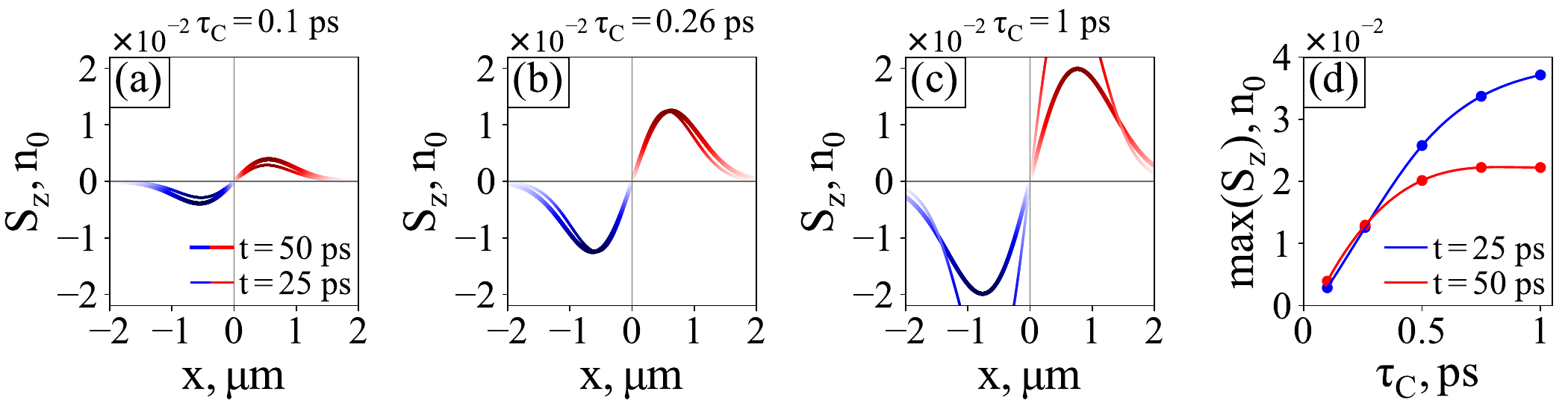}
    \caption{Spatial distribution of exciton spin density $S_z$ at different values of momentum relaxation time: (a) $\tau_{\rm C} = 0.1$ ps, (b) $\tau_{\rm C} = 0.26$ ps, (c) $\tau_{\rm C} = 1$ ps. 
    (d) Momentum relaxation time dependence of exciton spin density $S_z$ maximum.
   The increase in momentum relaxation time leads to an increase in the Hall current of intervalley excitons and reduces the spin relaxation time according to the Dyakonov-Perel mechanism, where spin relaxation time ($\tau_{\rm S}$) is proportional to the inverse of the momentum relaxation time ($\tau_{\rm C}$). The interplay of these two counteracting effects explains the saturation of the maximum value of $S_z$ shown in panel (d).
}
    \label{fig:S2}
\end{figure}

\begin{figure}[h!]
    \centering
    \includegraphics[width = \linewidth]{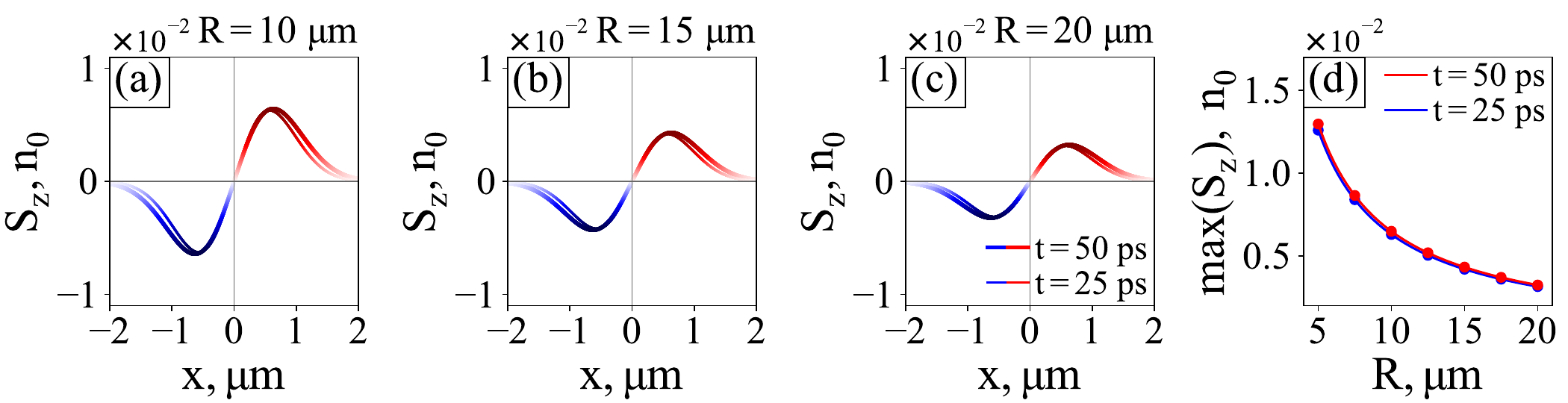}
    \caption{Spatial distribution of exciton spin density $S_z$ at different radii of the arc-shaped ribbon: (a) $R = 10$ $\rm \mu m$, (b) $R = 15$ $\rm \mu m$, (c) $R = 20$ $\rm \mu m$. (d) Dependence of exciton spin density $S_z$ maximum on the arc radius. It is evident that the maximum $S_z$ exhibits noticeable tunability with strain, characterized by $R$.}
    \label{fig:S3}
\end{figure}

Taking into account the influence of spin-relaxation effects, we present supporting figures \ref{fig:S1}, \ref{fig:S2}, and \ref{fig:S3} illustrating the dependence of $S_z$ on temperature ($T$), momentum relaxation time ($\tau_{\rm C}$), and the arc radius ($R$), respectively.


\end{document}